\documentclass[sigconf]{acmart}

\usepackage{libertine}
\usepackage{graphicx}
\usepackage{textcomp}
\usepackage{tikz}
\usepackage{caption}
\usepackage{subfig}
\newcommand{\circled}[1]{\tikz[baseline=(char.base)]{
            \node[shape=circle,fill=black,draw,inner sep=0.8pt] (char) {\fontsize{8}{8}\selectfont\textcolor{white}{#1}};}}
\newcommand{\FIXME}[1]{\color{red} #1 \color{black}}
\newcommand{\ADD}[1]{\color{blue}#1 \color{black}}
\newcommand{\DEL}[1]{\color{red}\sout{#1}\color{black}}
\newcommand{\seqnum}{SeqNum}
\newcommand{\eseqnum}{ESeqNum}
\newcommand{\acknum}{AckNum}
\newcommand{\mytitle}{ISN}
\newcommand{\mytitletwo}{RXL}

\usepackage{xcolor}
\usepackage{comment}
\usepackage[commandnameprefix=always]{changes}
\usepackage{algorithm}
\usepackage{algorithmicx, algpseudocode}
\usepackage{multirow}
\usepackage{tablefootnote}
\usepackage{threeparttable}
\usepackage{makecell}
\usepackage{array}
\usepackage{booktabs}
\usepackage{amsmath}

\DeclareGraphicsExtensions{.pdf, .png}

\AtBeginDocument{%
  \providecommand\BibTeX{{%
    Bib\TeX}}}

\setcopyright{none} 
\copyrightyear{2025}
\acmYear{2025}
\acmDOI{XXXXXXX.XXXXXXX}

\acmConference['SC 2025']{The IEEE/ACM International Conference for High Performance Computing, Networking, Storage, and Analysis}{Nov 16--21, 2025}{St. Louis, USA}
\acmISBN{979-8-3503-XXXX-X/18/06}

\begin{document}

%%
%% The "title" command has an optional parameter,
%% allowing the author to define a "short title" to be used in page headers.
\title{
\mytitletwo{}: Scaling Out Chip Interconnect Networks \\with Implicit Sequence Numbers}
  \title[Scaling Out Chip Interconnect Networks with Implicit Sequence Numbers]{Scaling Out Chip Interconnect Networks \\with Implicit Sequence Numbers}

% ACMart class 에서는 반드시 저자를 개별 선언하도록 되어 있음
%\author{Giyong Jung\textsuperscript{1}, Saeid Gorgin\textsuperscript{1}, John Kim\textsuperscript{2}, Jungrae %Kim\textsuperscript{1}}
%\affiliation{%
%  \textsuperscript{1}Dept. of Electrical and Computer Engineering, Sungkyunkwan University, Suwon, South Korea\\
%  \textsuperscript{2}School of Electrical Engineering, KAIST, Daejeon, South Korea
%}
%\email{jyk2498@skku.edu,gorgin_81@yahoo.com,jjk12@kaist.edu,dale40@skku.edu}

\author{Giyong Jung}
\affiliation{%
  \institution{Dept. of Electrical and Computer Engineering, Sungkyunkwan University}
  \city{Suwon}
  \country{South Korea}
}
\email{jyk2498@skku.edu}

\author{Saeid Gorgin}
\affiliation{%
  \institution{Dept. of Electrical and Computer Engineering, Sungkyunkwan University}
  \city{Suwon}
  \country{South Korea}
}
\email{gorgin_81@yahoo.com}

\author{John Kim}
\affiliation{%
  \institution{School of Electrical Engineering,\\ KAIST}
  \city{Daejeon}
  \country{South Korea}
}
\email{jjk12@kaist.edu}

\author{Jungrae Kim}
\affiliation{%
  \institution{Dept. of Electrical and Computer Engineering, Sungkyunkwan University}
  \city{Suwon}
  \country{South Korea}
}
\email{dale40@skku.edu}

\begin{abstract}
As AI models outpace the capabilities of single processors, interconnects across chips have become a critical enabler for scalable computing. These processors exchange massive amounts of data at cache-line granularity, prompting the adoption of new interconnect protocols like CXL, NVLink, and UALink, designed for high bandwidth and small payloads. However, the increasing transfer rates of these protocols heighten susceptibility to errors. While mechanisms like Cyclic Redundancy Check (CRC) and Forward Error Correction (FEC) are standard for reliable data transmission, scaling chip interconnects to multi-node configurations introduces new challenges, particularly in managing silently dropped flits in switching devices.

This paper introduces \emph{Implicit Sequence Number (\mytitle{})}, a novel mechanism that ensures precise flit drop detection and in-order delivery without adding header overhead. Additionally, we propose \emph{Reliability Extended Link (\mytitletwo{})}, an extension of CXL that incorporates \mytitle{} to support scalable, reliable multi-node interconnects while maintaining compatibility with the existing flit structure. By elevating CRC to a transport-layer mechanism for end-to-end data and sequence integrity, and relying on FEC for link-layer error correction and detection, \mytitletwo{} delivers robust reliability and scalability without compromising bandwidth efficiency.
\end{abstract}

\begin{comment}
Compute Express Link (CXL) is emerging as a key technology enabling high-speed communication between computing components in data centers.
However, as data rates increase, CXL becomes more vulnerable to transmission errors, while its small flit (flow control unit) sizes provide limited space for error protection mechanisms. Consequently, the current CXL protocol provides limited defense against flit drops, which can pose significant reliability challenges, particularly in configurations involving switching devices.

This paper introduces \emph{Implicit Sequence Number (ISN)}, a novel link layer enhancement designed to detect flit drops without requiring dedicated fields in the flit structure.
\FIXME{ISN can improve the reliability of CXL and other protocols while minimizing overhead.}
\end{comment}


%\received{20 February 2007}
%\received[revised]{12 March 2009}
%\received[accepted]{5 June 2009}

%%
%% This command processes the author and affiliation and title
%% information and builds the first part of the formatted document.
\maketitle

\section{Introduction}
\label{sec:Introduction}

% AI needs advanced chip interconnects
As Artificial Intelligence (AI) models continue to grow in size and complexity, their computational demands increasingly surpass the capabilities of individual processors ~\cite{GPU_Cluster2024,Evaluating_Emerging_AI_ML2024}. Training Large Language Models (LLMs) now requires extensive parallel processing across thousands of GPUs and CPUs~\cite{datacenter_scaleout1, datacenter_scaleout2}. For instance, Meta trained the 405-billion-parameter Llama 3.1 model using 16K NVIDIA H100 GPUs and 4K CPUs~\cite{llama3hom}. Similarly, state-of-the-art AI inference tasks now depend on the combined computational power of multiple GPUs~\cite{aminabadi2022deepspeed}.

% new protocols for advanced chip interconnects
To support such distributed computing at scale, high-throughput interconnects between CPUs, GPUs, and accelerators have become essential. This demand has driven the development of advanced interconnect protocols, such as NVIDIA’s NVLink~\cite{NVLink}, Compute Express Link (CXL)\cite{CXL_1_spec, CXL_2_spec, CXL_3_spec}, and Ultra Accelerator Link (UALink)\cite{McDowell2024}. These protocols deliver exceptional bandwidth through advanced \emph{physical layers} and operate on small data granularities (e.g., 64 bytes) at the \emph{link layer} to reduce latency and minimize overfetching. Additionally, these protocols incorporate cache coherence at the \emph{transaction layer}, allowing multiple chips to efficiently collaborate on shared data~\cite{CXLapplication1, CXLapplication2}.

% Scaling out interconnect infrastructure
Beyond protocol innovations, interconnect \emph{infrastructures} have also evolved to enable scalable deployments.
While traditional chip interconnects were initially designed for direct, point-to-point communication, modern systems now rely on scale-out architectures, where multiple processors communicate across shared networks. 
%\FIXME{Originally designed for direct, point-to-point communication between chips, modern chip interconnects now support scale-out architectures where numerous processors interact over shared networks.}
To facilitate this scalability, switching devices have become essential, enabling flexible scaling-out while preserving the high bandwidth of point-to-point links~\cite{datacenter_switch}.

% Challenge: transmission errors
However, as transfer rates rise and interconnect networks scale out, concerns over interconnect reliability are intensifying. For instance, during the 54-day training of Llama 3.1, Meta reported seven job interruptions attributed to NCCL (NVIDIA Collective Communications Library) watchdog time-outs~\cite{llama3hom}. Failures in NVLink often manifest as stalled load/store operations, with the system triggering a time-out upon detecting these stalls.
In a separate study, the Delta system~\cite{UIUC_Delta} experienced a Mean Time Between Errors (MTBE) of only 6.9 hours for NVLink errors across 848 NVIDIA Ampere GPUs, with 66\% of those errors resulting in job failures.

These incidents underscore the reliability challenges faced by modern interconnects. Furthermore, the problem is expected to worsen as signaling speeds escalate--from 50 GT/s in NVLink 3.0 (Delta) and 100 GT/s in NVLink 4.0 (Llama 3.1) to 200 GT/s in NVLink 5.0. Such aggressive scaling tightens timing margins and amplifies signal integrity issues, making transmission errors both more likely and more disruptive, especially in large-scale deployments.

% Current solution
To mitigate the rising physical-layer error rates, modern protocols incorporate robust link-layer error recovery mechanisms. For example, CXL 3.0 doubles the transfer rate of its predecessor to 64 GT/s by relaxing the acceptable Bit Error Rate (BER) to $10^{-6}$--six orders of magnitude higher than CXL 2.0’s $10^{-12}$. To maintain reliable communication, CXL 3.0 introduces an 8-byte Cyclic Redundancy Check (CRC) and 6-byte Forward Error Correction (FEC) per 256-byte flit~\cite{CXL_2_spec, CXL_3_spec}. CRC detects flit corruption for retransmission, while FEC corrects many errors on-the-fly, reducing the reliance on retransmission and ensuring robust performance.

% Issues of the current solutions
While CRC and FEC provide effective protection in direct links, these mechanisms fall short in more complex, switched environments. 
In direct connections, CRC and FEC rely on the flit successfully reaching its destination to trigger error recovery~\cite{PCIe_6_spec}. 
However, in switched environments, intermediate devices often discard erroneous flits without notifying the final destination~\cite{PCIeswitch1, PCIe_switch2, PCIe_switch3, Ethernet_switch1, Ethernet_switch2}, leading to silent flit drops. These drops disrupt sequence alignment and data integrity at the endpoint, particularly in systems lacking mechanisms to track dropped flits~\cite{seqnumattack}.

Traditional networking protocols, such as the \emph{Transmission Control Protocol (TCP)}, address such challenges by incorporating two dedicated fields in their headers for sequence tracking: the \emph{sequence number (\seqnum{})}, which indicates the order of the current packet, and the \emph{acknowledgment number (\acknum{})}, which represents the last correctly received packet in the reverse direction (i.e., via ACK piggybacking)~\cite{tanenbaum2021computernetworks}. Together, these fields allow the sender and receiver to detect and recover from missing or out-of-order packets, ensuring reliable and ordered communication even in complex network environments.

However, this approach introduces substantial header overhead, such as the 4-byte \seqnum{} and 4-byte \acknum{} fields in TCP. While such overhead is manageable for networking payloads, which are often a few kilobytes in size, it is impractical for chip interconnect protocols with much smaller payloads (e.g., 64 bytes), where minimizing header size is critical. For example, CXL's 256-byte flits dedicate only a two-byte header to sequence tracking and other control fields.

As a result, CXL’s 256-byte flits allocate a single 10-bit field for sequence tracking, which serves dual purposes as either a \seqnum{} or an \acknum{}~\cite{CXL_3_spec}. While this approach minimizes header overhead, it introduces inherent limitations. A single field cannot simultaneously manage sequencing and acknowledgment effectively. For example, when a flit uses the field to carry an \acknum{}, the receiver loses visibility of the flit’s \seqnum{}, making it impossible to confirm whether the flit is part of the expected sequence (Section \ref{sec:prior work:cxl}). This lack of sequence tracking creates critical vulnerabilities in multi-node environments where switching devices are prone to dropping flits (Section \ref{sec:eval:reliability}).

\begin{comment}
For example, CXL employs 68-byte flits with a 2-byte CRC to detect flit corruption but omits additional fields for identifying flit drops. Similarly, CXL’s 256-byte flits include only a single 10-bit field for sequence tracking, serving dual roles as either a \seqnum{} or an \acknum{}. While this design reduces header size, it introduces limitations. A single field cannot reliably manage both sequencing and acknowledgment. For instance, when a flit carries an \acknum{}, the receiver cannot verify whether it is the next flit in the sequence, creating potential gaps in sequence integrity. These limitations pose significant challenges in multi-node environments with switching devices, where dropped flits are more likely to occur.
\end{comment}

This paper introduces a novel concept called \emph{Implicit Sequence Number (\mytitle{})}, designed to address the challenges of achieving accurate sequence tracking while maintaining minimal flitization overhead. \mytitle{} eliminates the need for explicitly transmitting a \seqnum{} in flit headers by implicitly embedding the sequence information within the CRC checksum. This approach leverages the inherent redundancy in error recovery mechanisms, as both CRC violations and \seqnum{} mismatches ultimately trigger the same response: \emph{retry}.

\begin{figure}[t] % t: top, b: bottom, h: here
    \centering
    \subfloat[An example of flit processing with explicit sequence numbers. The flit header includes a sequence number.]{
        \includegraphics[width=.8\linewidth]{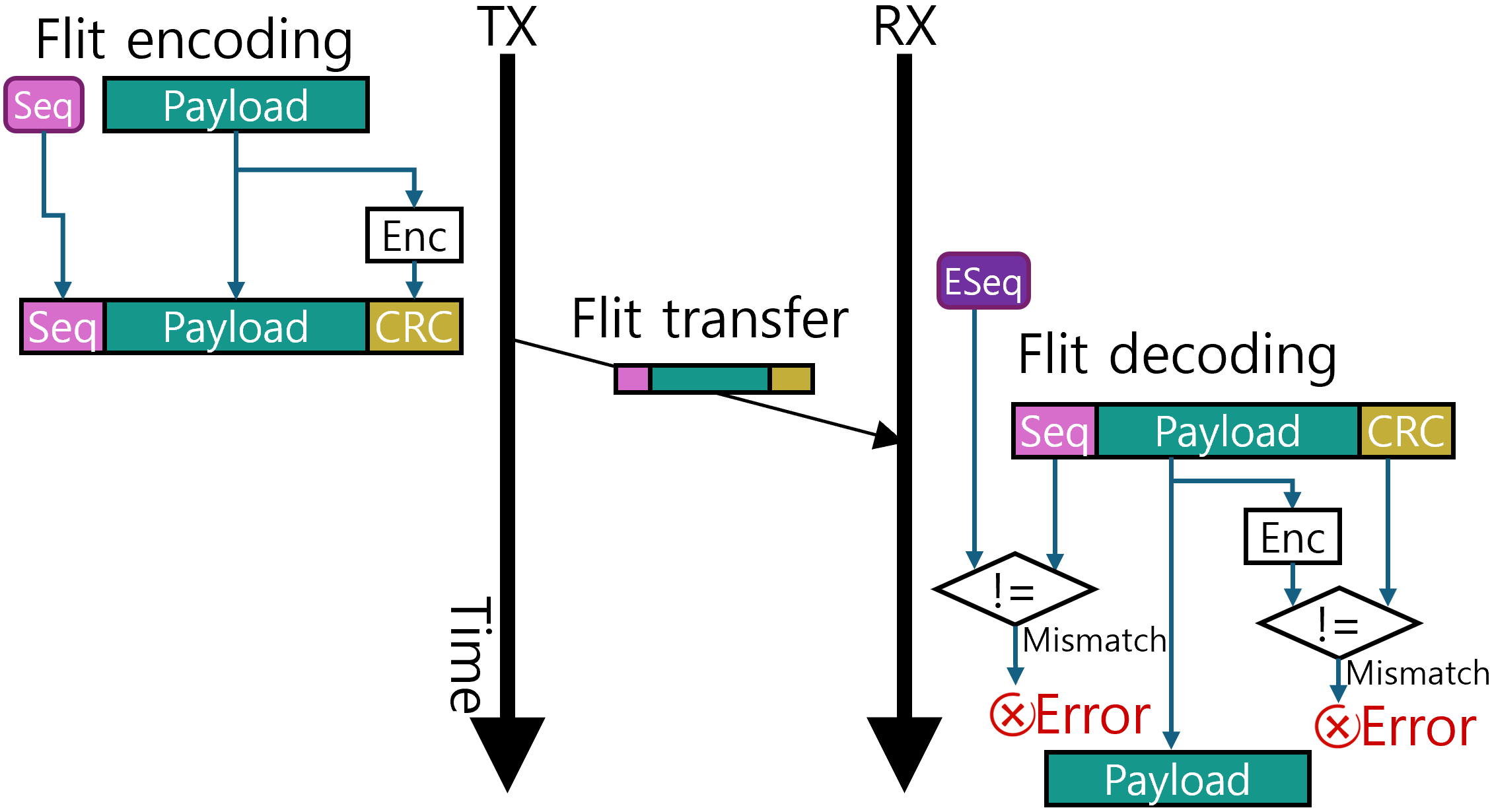}
        \label{fig:ESN}
        \vspace{-8pt}
    }\\
    \subfloat[An example of flit processing with \emph{implicit sequence numbers}. The header has no sequence number but the CRC is generated based on both the payload \emph{and} sequence number.]{ \includegraphics[width=.8\linewidth]{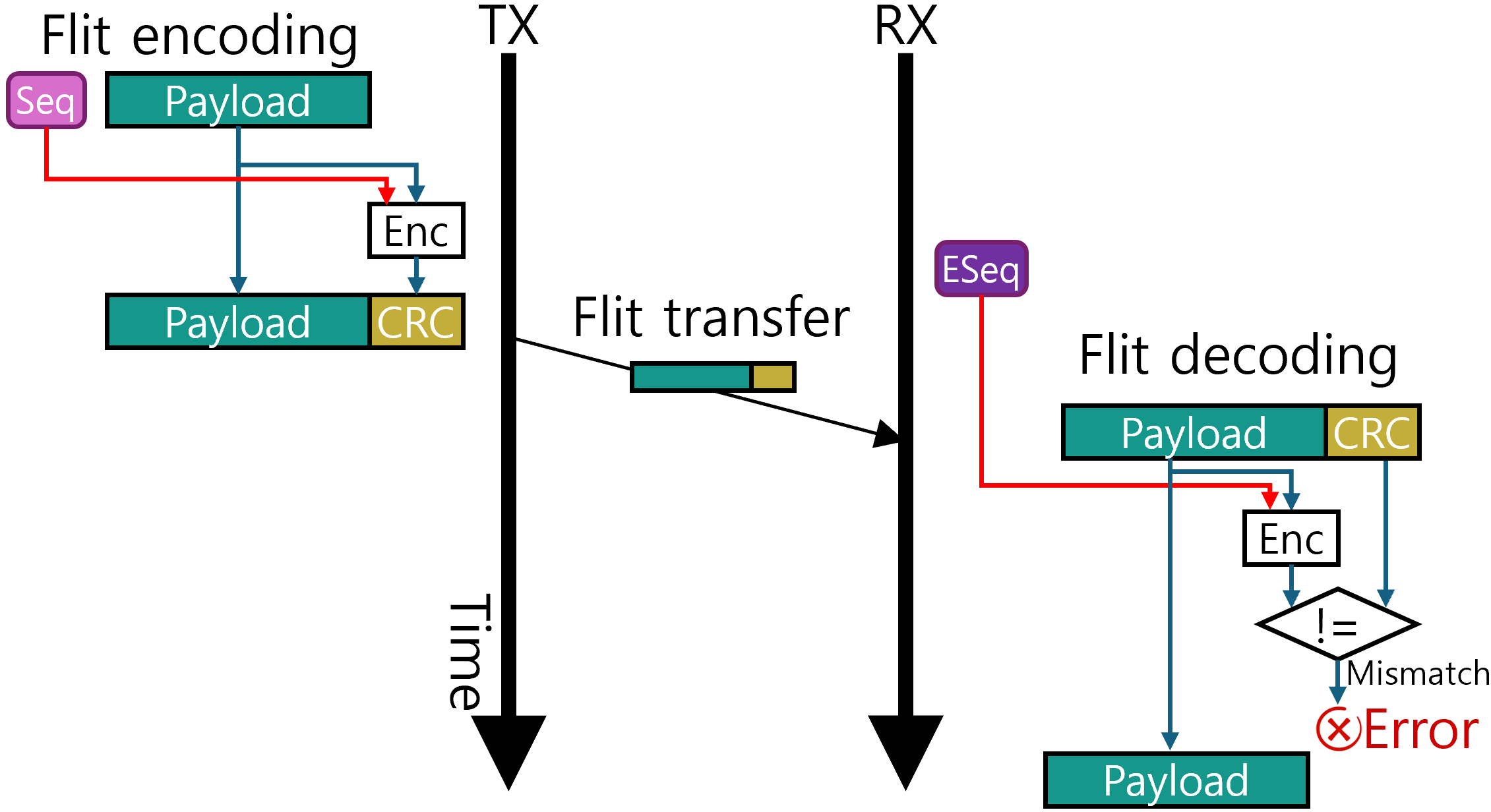}
        \label{fig:ISN}
        \vspace{-8pt}
    }\\
    \vspace{-5pt}
    \caption{Comparison between explicit sequence numbers (used in prior work) and implicit sequence numbers (\mytitle{}).}
    \label{fig:intro}
    \vspace{-15pt}
\end{figure}

Fig. \ref{fig:intro} illustrates the difference between traditional explicit sequence numbers and \mytitle{}. In \mytitle{}, the CRC is encoded using both the payload and the \seqnum{}, effectively embedding the sequence number into the CRC checksum itself. At the receiver, CRC decoding incorporates the received payload and the \emph{expected sequence number (\eseqnum{})}. Any mismatch--either in the payload or between \seqnum{} and \eseqnum{}--triggers a CRC error, enabling the detection of missing flits through CRC results alone.

By eliminating the need for explicit sequence number fields, \mytitle{} offers a lightweight and practical solution that is compatible with a broad range of interconnect protocols. Many open standards, such as PCIe~\cite{PCIe_6_spec} and UCIe~\cite{UCIe_spec}, already employ compact headers in which a single field is multiplexed as either a \seqnum{} or \acknum{}, thereby inheriting similar limitations. Although proprietary protocols like NVLink provide limited public documentation, they are likely to adopt comparable strategies for minimizing header overhead. In this work, we apply \mytitle{} to the CXL protocol and demonstrate how implicit sequence tracking can substantially improve reliability—particularly in scale-out architectures that rely on switching devices.

The enhanced protocol, referred to as \emph{\mytitletwo{} (Reliability Extended Link)}, integrates \mytitle{} into the transport layer of CXL, extending reliability across the entire interconnect network. In CXL, CRC and FEC are confined to the link layer, providing error detection and correction for individual links. In contrast, \mytitletwo{} shifts the CRC functionality to the transport layer, enabling end-to-end protection for both data integrity and sequence integrity. \mytitletwo{} generates the CRC using \seqnum{} at the originator and decoded at the final destination using the \eseqnum{}. This shift eliminates the need for switching devices to manage sequence numbers for the modified CRC decoding and enables end-to-end data protection against link-level errors and internal errors within switches. Meanwhile, FEC continues to operate at the link layer in \mytitletwo{}, providing robust error correction and detection to minimize unnecessary transfers of erroneous flits. This dual-layer approach ensures reliable communication across multi-node systems, making \mytitletwo{} a scalable and efficient interconnect solution for modern distributed computing environments.

The major contributions of this paper are as follows:
\begin{itemize}
    \item We identify critical limitations of current chip interconnect protocols in handling silently dropped flits, which hinder its ability to scale out effectively. To the best of our knowledge, this is the first work to address protocol-level reliability challenges in chip interconnects.
    \item We propose \emph{Implicit Sequence Number (\mytitle{})}, a novel mechanism that integrates sequence tracking into the CRC, eliminating the need for explicit sequence numbers in headers.
    \item We introduce \emph{\mytitletwo{}}, an enhancement of the CXL protocol that implements \mytitle{}. \mytitletwo{} enables robust detection of sequence misalignments and silent flit drops while maintaining compatibility with existing flit structures.
\end{itemize}

\section{Background}
\label{sec:background}

\subsection{Chip Interconnect Errors}
\label{sec:background:interconnect_errors}

Errors on chip interconnects have historically been underexplored by the computer architecture community. However, field studies have suggested that interconnect reliability can become a critical bottleneck in large-scale systems~\cite{gupta2017failures, di2019mira, llama3hom}.
For instance, a study on the Jaguar XT5 supercomputer~\cite{gupta2017failures} reported that approximately 10\% of application interruptions were related to its SeaStar2+ communication processors~\cite{abts2011cray}. Similarly, the IBM Mira supercomputer has reported that 21\% of fatal events are related to its Blue Gene/Q link modules~\cite{di2019mira}.

More recently, Meta conducted a root-cause analysis of job interruptions during the 54-day training of its Llama 3.1 model with 405B parameters~\cite{llama3hom}. This system utilized 16K NVIDIA H100 GPUs, with eight GPUs per server interconnected via NVLink 4.0 (100 GT/s). Among 322 job interruptions caused by hardware failures, 7 cases (2.2\%) were attributed to NCCL watchdog timeouts, a potential indicator of NVLink issues.

A separate field study on the Delta system~\cite{UIUC_Delta} observed that 4.7\% of GPU errors were due to NVLink errors (XID=74) in the 206-node system comprising 848 NVIDIA Ampere GPUs. These errors translated to a MTBE of 1,415 node hours or just 6.9 hours at the system level. Alarmingly, 66\% of the NVLink errors led to job failures, and 14\% affected two or more GPUs, despite the use of mitigation strategies such as memory error containment.
In terms of job interruptions, NVLink errors accounted for 3.7\% of all cases. However, after excluding a small number of faulty HBMs responsible for 90\% of uncontained memory errors, the contribution of NVLink errors rose significantly—accounting for 10.2\% of the remaining job interruptions. This finding underscores the growing impact of interconnect errors in modern GPU clusters, particularly as other sources of failure become better contained.

These reliability concerns are expected to worsen as interconnect bandwidth continues to scale aggressively.
For instance, CXL 3.0, operating at 64 GT/s, has relaxed its BER tolerance to $10^{-6}$. Similarly, the UALink protocol~\cite{UALink_spec}, which is based on the physical layer of 200 GT/s Ethernet (IEEE 802.3dj), could experience a BER as high as $2.4 \times 10^{-4}$\cite{ethernet_standard}. BER requirements of proprietary solutions like NVIDIA's NVLink 5.0 (200 GT/s)\cite{nvidia2024nvlink} and Google's ICI (Inter Chip Interconnect, 400 GT/s)~\cite{semianalysis2023ici} are not publicly available, but they are highly likely to face comparable or even greater error vulnerabilities due to their aggressive data rates.
These trends underline the urgency of addressing interconnect reliability in future large-scale systems.

%Although these interconnect-related failures were fewer than those caused by GPUs (46.0\%) and HBMs (22.4\%), which are known to be less reliable, they equaled the combined failure frequencies of CPU (0.6\%), CPU memory (0.6\%), and SSD (0.9\%). These findings underline the increasing prominence of chip interconnect errors, even in cutting-edge systems.

\begin{comment}
\begin{table}[h!]
    \centering
    \tiny
    \caption{A comparison of chip interconnect protocols.}
    \resizebox{\columnwidth}{!}{
    \begin{tabular}{|c|c|c|c|c|}
        \hline
        Protocols   & \makecell{Max.\\rate}   & \makecell{Max.\\BER}   & \makecell{Flit\\size} & \makecell{CRC/FEC\\size}  \\
        \hline
        CXL 2.0     & \makecell{32 GT/s}                            & $10^{-12}$            & 68B                   & 16-bit CRC   \\
        \hline
        CXL 3.0     & \makecell{64 GT/s}                            & $10^{-6}$             & 256B                  & \makecell{64-bit CRC\\48-bit FEC} \\
        \hline
        
        NVLink4  & \makecell{100 GT/s} & ? & ? & \makecell{CRC\\FEC} \\
        \hline
        NVLink5  & \makecell{200 GT/s} & ? & ? & \makecell{CRC\\FEC} \\
        \hline
    \end{tabular}
     }
\end{table}
\end{comment}

\subsection{Compute Express Link (CXL)}
\label{sec:background:cxl}

CXL is an open-standard, cache-coherent interconnect designed to enable low-latency, high-bandwidth communication among CPUs, memory, accelerators, and peripheral devices.
Built on top of the PCIe \emph{physical layer}, CXL introduces three distinct protocols at the \emph{link} and \emph{transaction} layers: \texttt{CXL.io}, \texttt{CXL.cache}, and \texttt{CXL.mem}~\cite{CXL_3_spec}. While CXL.io maintains compatibility with PCIe for basic I/O operations, CXL.cache and CXL.mem enable low-latency, high-bandwidth memory access with cache coherence. 
According to a recent study, CXL can deliver up to $5.7\times$ higher data throughput in multi-node systems compared to PCIe~\cite{TAROT}, underscoring its advantage in supporting data-intensive applications in data centers~\cite{Sharma2023}.

CXL achieves high bandwidth through advanced physical layer techniques. CXL 3.0 employs PAM4 (Pulse Amplitude Modulation with four levels), which transmits 2 bits per symbol at a signaling rate of 64 GT/s~\cite{intel2019pam4,PAM4_2020}. However, this modulation imposes a tight 31ps timing window per symbol, making signal integrity a significant challenge. To mitigate inter-symbol interference (ISI), CXL incorporates Decision Feedback Equalization (DFE), which adjusts signal levels based on previously received symbols~\cite{DFE2024}. While DFE effectively mitigates ISI, an incorrect decision can propagate errors across subsequent symbols~\cite{DFE_error2009}.

Reflecting these challenges, CXL 3.0 relaxed its bit error rate (BER) tolerance to $10^{-6}$--a sharp increase from CXL 2.0's $10^{-12}$. As a result, a full-speed $\times$16 CXL 3.0 link (1 Tb/s) may experience up to one million bit errors per second. 
Furthermore, these \emph{first bit errors} can further propagate through the DFE, manifesting as burst errors at the link layer.
These conditions necessitate robust error detection and correction mechanisms above the physical layer to ensure end-to-end reliability.

The CXL link layer ensures reliable data transfer by managing flow control and error handling. To accommodate high data rates and small block sizes (e.g., transferring 256 bytes over a $\times16$ CXL 3.0 link takes just 2 ns), data is transmitted in fixed-size flow control units (flits) rather than variable-size packets. CXL 3.0 defines two flit sizes: 68B and 256B ~\cite{CXL_3_spec}. In the full-speed mode (64 GT/s), 256B flits are employed to improve reliability with additional headers. At reduced speeds, 68B flits can be used to minimize latency and reduce overfetching. 

At the transaction layer, CXL supports cache-coherent access between hosts and devices, based on the MESI (Modified, Exclusive, Shared, Invalid) protocol ~\cite{cxl_intro}. A transaction involves three \emph{messages}: a request, a response, and data. For example, a device may issue a RdCurr request to retrieve the latest cache line version from the host. The host processes this request, updates its cache hierarchy, sends a response, and then transfers the requested data. To improve link efficiency, the link layer can pack tens of transaction messages into a single 256B flit.

\begin{comment}
The CXL link layer is responsible for reliable data transfer by handling framing, flow control, and error management. Data is transmitted in fixed-size flow control units (flits), which maintain consistency across endpoints and switching devices. CXL employs 68B flits in CXL 1.0, 2.0, and degraded modes of CXL 3.0 to transfer 64B of payload. With the introduction of CXL 3.0, 256B flits have been adopted to transfer 240B payloads, allowing for enhanced reliability, more comprehensive headers, and support for complex topologies.
\end{comment}

\subsection{CXL Flit Errors}
\label{sec:background:flit_errors}

Bit errors in CXL transmissions manifest as \emph{flit corruption} or \emph{flit drops}. Flit corruption occurs when an erroneous flit reaches the endpoint. At the endpoint, the error can be detected using CRC, and a retry is initiated to recover the corrupted data.

In contrast, flit drops occur when an erroneous flit is discarded by an intermediate switching device. The CXL standard does not explicitly define how switches should handle corrupted flits, and no publicly available data sheets provide clarification. However, datasheets for existing PCIe and Ethernet switches~\cite{PCIeswitch1, PCIe_switch2, PCIe_switch3, Ethernet_switch1, Ethernet_switch2} suggest that switching devices typically discard packets upon detecting errors, which is practical because forwarding corrupted flits risks misrouting caused by undetectable address corruption~\cite{Sharma2024_FEC_CRC}.
%From the endpoint's perspective, such discarded flits are effectively lost, posing a challenge for ensuring reliable communication in scaled-out interconnects.

Flit corruption and drops pose significant threats to system reliability. These errors can propagate to higher layers, jeopardizing computation accuracy and overall system stability. Cache-coherent protocols are particularly vulnerable because they rely on the strict ordering of requests, responses, and data to maintain consistency. Any disruption caused by flit drops can lead to cache inconsistencies, erroneous data sharing, and unpredictable behavior.

The risks are amplified by CXL’s ability to pack multiple transaction messages into a single flit. For example, a 256B flit can carry up to 44 messages (request, data-header, or response) per 128B group. Losing such a flit could disrupt all associated transactions, amplifying the impact on system stability and data integrity.
%As CXL scales to multi-node configurations with switching devices, addressing these challenges becomes critical to ensure robust and reliable communication ~\cite{CXL_multiNode2024}.

\subsection{TCP Reliability Mechanisms}
\label{sec:background:tcp}

CXL relies on its link layer alone for reliable transmission over individual links.
In contrast, scaled-out network protocol stacks like TCP/IP incorporate a transport layer on top of a link layer to ensure accurate, ordered, and end-to-end data delivery across networks~\cite{rewaskar2006empirical}. This section reviews the reliability mechanisms of TCP, a transport layer protocol designed to handle the diverse and error-prone nature of the Internet. \mytitletwo{} adapts some of these ideas to enhance reliability in scaled-out interconnect networks while preserving the existing CXL flit structure.
% and avoiding additional header field.

\begin{figure}[t] % t: top, b: bottom, h: here
    \centering \includegraphics[width=0.75\columnwidth]{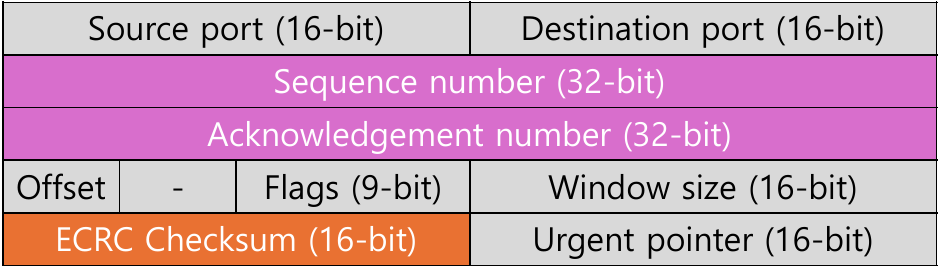}
    \vspace{-10pt}
    \caption{The TCP header format (20B).}
    \label{fig:tcp-header}
    \vspace{-15pt}
\end{figure}

The TCP header format (Fig. \ref{fig:tcp-header}) includes three key fields for reliability: the 32-bit \emph{sequence number (\seqnum{})}, the 32-bit \emph{acknowledgment number (\acknum{})}, and the 16-bit \emph{End-to-End CRC (ECRC)}.

\subsubsection{Sequence Number}

This field ensures in-order delivery of data by uniquely identifying the position of each segment in the data stream. During connection setup, the sender and receiver exchange 32-bit initial sequence numbers to synchronize communication. Then the sender includes the \emph{sequence number} in the TCP header of each transmitted segment and increments it for subsequent segments. The receiver tracks an \emph{expected sequence number}, which allows it to detect missing or out-of-order segments, enabling detection of packet drops and reassembly of the data stream.

%TCP assigns a 32-bit sequence number to each segment to ensure correct data ordering. These sequence numbers allow the receiver to reassemble data accurately, even if segments arrive out of order. They also enable the receiver to detect missing segments and signal potential packet drops.

\subsubsection{Acknowledgment Number}

Acknowledgment numbers complement sequence numbers by providing feedback on received segments. If a segment is missing or fails CRC validation, the receiver sends a NACK (Negative ACKnowledgment) containing the last correctly received sequence number, prompting the sender to retransmit the missing data. Upon successful receipt, the receiver sends an ACK (ACKnowledgment) to free the sender’s buffer for acknowledged segments.

ACK messages can generate significant traffic in the reverse direction. To reduce the reverse-direction traffic, TCP employs two mechanisms: \emph{ACK coalescing} and \emph{ACK piggybacking}. Coalesced ACKs aggregate acknowledgments to reduce the frequency of individual ACKs, while piggybacking attaches an acknowledgment to the next outgoing data packet.
To support piggybacking, TCP headers include an additional sequence number: a 32-bit acknowledgment number (\acknum{}), allowing both data transmission and acknowledgment in a single packet.

\begin{comment}
\ADD{
\subsubsection{TCP Timeouts}
TCP employs timeouts as a critical mechanism to ensure reliable data delivery. When a packet is sent, a retransmission timer starts, and if an acknowledgment (ACK) isn't received before the timer expires, TCP assumes the packet is lost or corrupted and retransmits it. The timeout value, or Retransmission Timeout (RTO), is dynamically calculated based on the Smoothed Round-Trip Time (SRTT) and its variation (RTTVAR), adapting to network conditions. To prevent overloading the network during congestion, TCP employs exponential backoff, doubling the RTO for each failed retransmission. Additionally, mechanisms like Fast Retransmit can trigger retransmissions before a timeout occurs, minimizing delays.
}
\end{comment}

\subsubsection{End-to-End CRC}

Although the link layer utilizes CRC to allow reliable transmission within a link, errors outside individual links (e.g., within switching devices) are not protected by the \emph{Link CRC (LCRC)}. To fill this reliability gap, TCP utilizes a 16-bit checksum as End-to-end CRC (ECRC). The sender generates the checksum and embeds it in the segment header. At the endpoint, the receiver recalculates the checksum and compares it to the transmitted value to ensure end-to-end data integrity.

While TCP’s mechanisms provide robust reliability, they incur significant header overhead. For example, the TCP header size is 20B, and when combined with IPv6 (40B) and Ethernet (14B) headers, the total overhead reaches 74B. This overhead is typically amortized by large payloads (e.g., 1 KB) of network traffic. However, it is less suited for small data units like those used in chip-to-chip interconnects.

\subsection{Shortened Reed Solomon Codes}
\label{sec:background:shortened}

Wired communication systems often employ Reed-Solomon (RS) codes for FEC due to their robust and efficient error-correction capability ~\cite{LinError2004}. These codes can correct up to $t$ symbol errors within a codeword by utilizing $2t$ redundant symbols. The maximum codeword size is $2^m-1$ symbols, where $m$ is the size of each symbol in bits. For instance, with 8-bit symbols, RS codes can provide \emph{Single Symbol Correction (SSC)} for a 255-symbol codeword using just 2 redundant symbols ~\cite{computing1986fault}.

In practical applications such as CXL, however, the data block sizes are often smaller than the maximum codeword size. To accommodate this, RS codes are typically \emph{shortened}. During the encoding process, unused symbol positions in the codeword are padded with zeroes to match the actual data block size. At the decoding stage, any correction attempt targeting these constant zero-padded positions is flagged as invalid, enabling the decoder to detect some errors beyond the correction threshold.

This extra detection capability allows FEC to function as a robust first line of defense against transmission errors.
In CXL, a 256B flit is divided into three sub-blocks of 85, 85, and 86 bytes. Each sub-block is protected by 8-bit symbol RS codes, which provide SSC with 2 redundancy bytes, as shown in Fig. \ref{fig:256B_flit}. These shortened codes can detect approximately two-thirds of uncorrectable errors due to the presence of 170 unused positions in the 255-symbol codeword. As a result, the FEC with 3-way interleaved SSC can correct up to 3-symbol burst errors and detect 2/3 of 4-symbol burst errors, 8/9 of 5-symbol burst errors, and 26/27 of 6-symbol burst errors or larger, as it requires all three sub-blocks to miscorrect for such burst errors to go undetected. 

\mytitletwo{} leverages these detection capabilities of FEC to restructure the CXL protocol stack. It exclusively relies on FEC for local error detection and correction at the link layer, while elevating CRC to the transport layer for end-to-end validation of both data and sequence integrity at the endpoints (Section \ref{sec:i-seq}).

\section{PriorWork}
\label{sec:prior work:other}

The concept of leveraging data protection mechanisms to encode auxiliary metadata has been explored in several prior works, including Implicit-Storing~\cite{sazeides2013is}, All-Inclusive ECC~\cite{kim2016allinclusive}, and Implicit Memory Tagging~\cite{sullivan2023imt}. These approaches utilize ECC redundancy to protect address signals or to embed metadata such as memory tags without requiring additional storage overhead.

In contrast, this work targets the unique reliability challenges of high-speed, multi-node chip interconnects. Specifically, it addresses sequence ordering violations caused by flit drops in scaled-out interconnects, a critical issue overlooked in current protocols. By embedding sequence tracking into CRC via \mytitle{}, this approach ensures robust sequence validation and end-to-end data integrity for chip interconnects, offering a scalable solution to the growing challenges posed by error-prone interconnect technologies.

Recent works on memory ECC have leveraged the underutilized syndromes of shortened RS codes to enhance their correction capabilities~\cite{UNITY, SELCC, SECBADAEC, DAECC}. In contrast, \mytitle{} takes advantage of the partial detection capabilities of shortened RS codes to enable early error detection based on FEC.

\section{Motivation}
\label{sec:prior work}

\begin{figure}[t] % t: top, b: bottom, h: here
    \centering
    \includegraphics[scale=0.45]{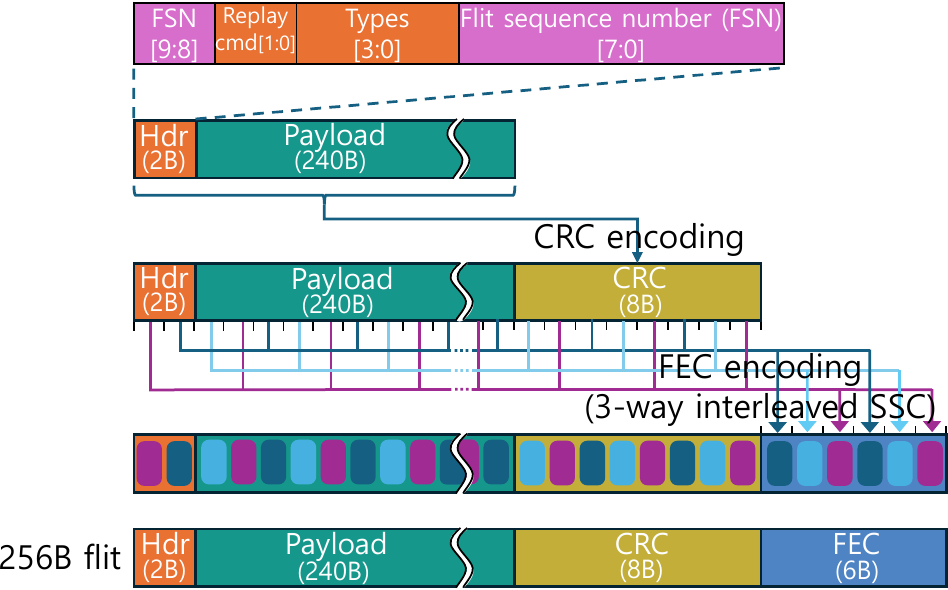}
    \vspace{-10pt}
    \caption{The structure and encoding of CXL 256B flits.}
    \label{fig:256B_flit}
    \vspace{-15pt}
\end{figure}

\begin{comment}
\begin{figure}[t] % t: top, b: bottom, h: here
    \centering
    \includegraphics[scale=0.45]{Figures/256B_flit.pdf}
    \vspace{-10pt}
    \caption{The structure and encoding of CXL 256B flits.}
    \label{fig:256B_flit}
    \vspace{-15pt}
\end{figure}
\end{comment}

This section examines the reliability mechanisms employed in CXL 3.0 and their weakness. The discussion primarily centers on the 256B flit structure, as 68B flits are limited to lower-speed modes and are unsuitable for high-performance configurations.

\subsection{Analysis of CXL Reliability}
\label{sec:prior work:cxl}
CXL 3.0 incorporates several reliability mechanisms to handle its high frequency of bit errors (up to $10^{-6}$ BER).
Each 256B flit comprises a 2B header, a 240B payload, an 8B CRC, and a 6B FEC (Fig. \ref{fig:256B_flit}).
The 8B CRC is generated from the 2B header and 240B payload, providing robust detection capabilities. It can detect up to four random bit errors and burst errors up to 64 bits long with complete reliability. For more severe errors, it achieves an extremely high detection probability of $1 - 2^{-64}$. This ensures that even rare multi-bit errors are identified with near certainty.

The FEC splits the 250B combined header, payload, and CRC into 83B, 83B, and 84B sub-blocks and appends 2B redundancy to each sub-block using 8-bit symbol RS codes. These redundant symbols enable SSC within each sub-block, while the 3-way interleaving mechanism allows the FEC to correct up to three-symbol burst errors. According to the PCIe 6.0 standard, the flit error rate after FEC must not exceed $3\times10^{-5}$~\cite{PCIe_6_spec}.

This combination of CRC and FEC provides robust error detection and correction, significantly enhancing data integrity. 
Uncorrectable flits—those with errors spanning multiple symbols within a sub-block—are detected with moderate probability by the FEC (Section \ref{sec:background:shortened}) and with exceedingly high probability by the CRC, which boasts an undetected flit error rate below $1.6 \times 10^{-24}$ (Section \ref{sec:eval:reliability}).
To achieve this robustness, however, CXL 3.0 allocates 5.5\% of the flit size (14 bytes) to redundancy.

While CRC and FEC provide robust protection against data corruption, they are ineffective against silent flit drops, as their operations are confined to received data. To address this limitation, CXL 3.0 incorporates a 10-bit \emph{Flit Sequence Number (FSN)} and a 2-bit \emph{ReplayCmd} field within the 2B header ~\cite{CXL_3_spec} (Fig. \ref{fig:256B_flit}). The FSN serves multiple purposes based on the ReplayCmd setting:
\begin{itemize}
    \item $ReplayCmd = 0$: The FSN contains the explicit sequence number of the current flit.
    \item $ReplayCmd = 1$: The FSN carries the acknowledgment sequence number (\acknum{}).
    \item $ReplayCmd = 2$: The FSN specifies the last valid received \seqnum{} for NACK with go-back-N retry.
    \item $ReplayCmd = 3$: The FSN specifies the last valid received \seqnum{} for NACK for single flit retry.
\end{itemize}

\begin{figure}[t] % t: top, b: bottom, h: here
    \centering
    \includegraphics[scale=0.46]{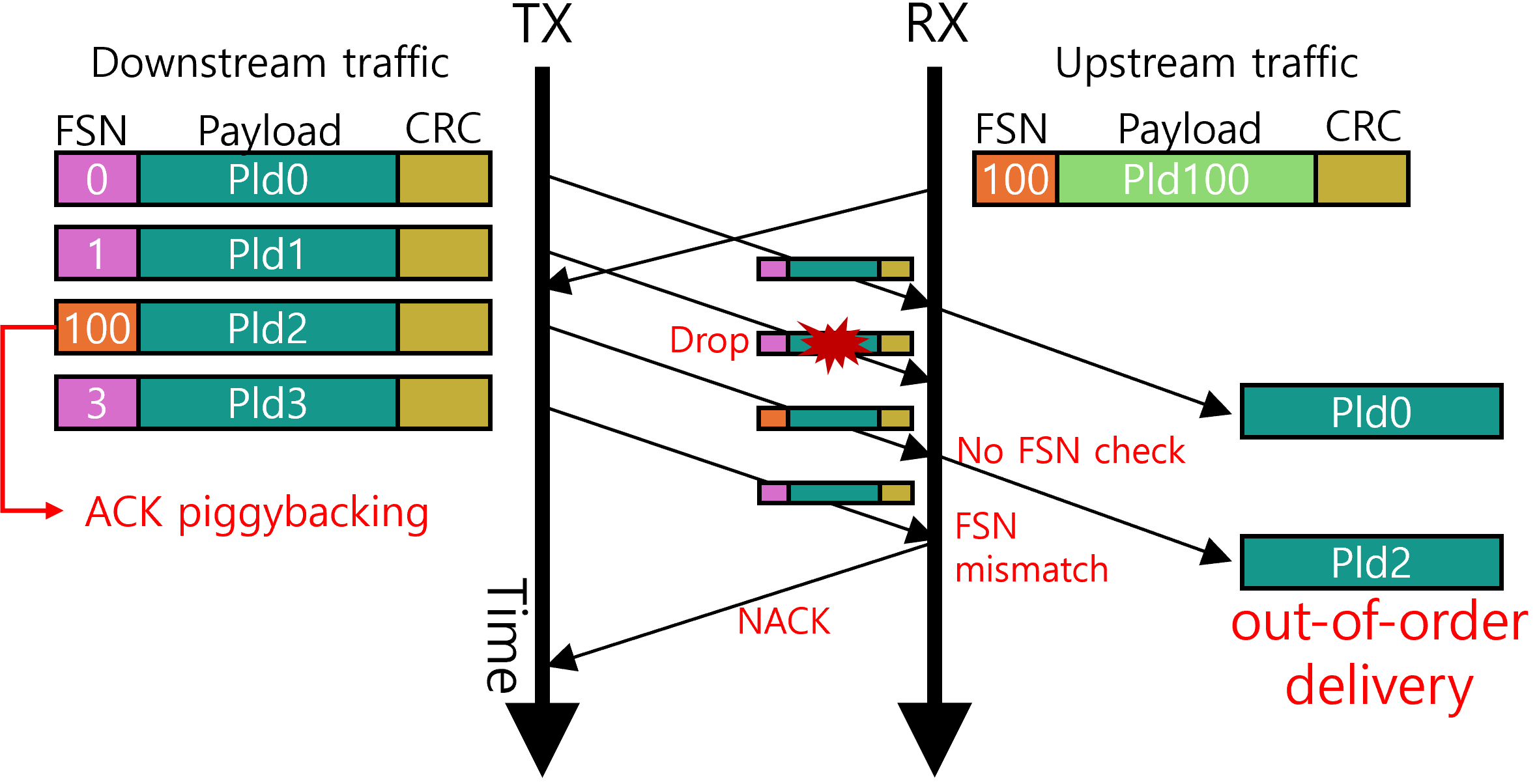}
    \vspace{-10pt}
    \caption{An example of CXL's failure on handling flit drops at the link layer.}
    \label{fig:256B-failure}
    \vspace{-15pt}
\end{figure}

While this multi-purpose FSN design reduces header size, it introduces reliability trade-offs. Fig. \ref{fig:256B-failure} demonstrates a potential failure scenario in CXL's handling of flit drops. In this example, a host transmits downstream flits numbered \#0, \#1, \#2, and \#3, while the device sends an upstream flit numbered \#100. To acknowledge the upstream flit, flit \#2 carries an \acknum{} (100) in its FSN field instead of its own sequence number (2).

Suppose flit \#1 is dropped due to an error. The device should ideally halt the processing of subsequent flits to maintain the ordering. However, because flit \#2 does not include its own \seqnum{}, the device can only conduct data integrity checks (using CRC and FEC) but cannot verify sequence integrity.
Upon passing the data integrity check, the device forwards flit \#2 to application layers, despite the missing flit \#1.

Later, when flit \#3 arrives, the device can detect the missing flit by comparing the \seqnum{} (3) with the \eseqnum{} (2). However, by this time, the system has already forwarded the incorrectly ordered flit sequence to the application layer, rendering the late detection ineffective.
This premature forwarding can potentially disrupt a large number of messages packed within the missing flit (up to 44 messages per 128B group), which could lead to unpredictable behaviors and inconsistencies across caches (Section \ref{sec:prior work:cxl2}).

This reliability hole due to incomplete sequence tracking can be exacerbated in multi-level switching environments. Although the switches can correct most erroneous flits using FEC, the remaining flits can result in intermediate flit drops. As the number of switching levels increases, these uncorrectable flit drops can accumulate, causing a proportional rise in flit drops at the endpoint.

\subsection{Impact of CXL Vulnerability}
\label{sec:prior work:cxl2}

\begin{figure}[t] % t: top, b: bottom, h: here
    \centering
    \subfloat[A failure due to duplicate requests.]{
        \includegraphics[scale=0.48]{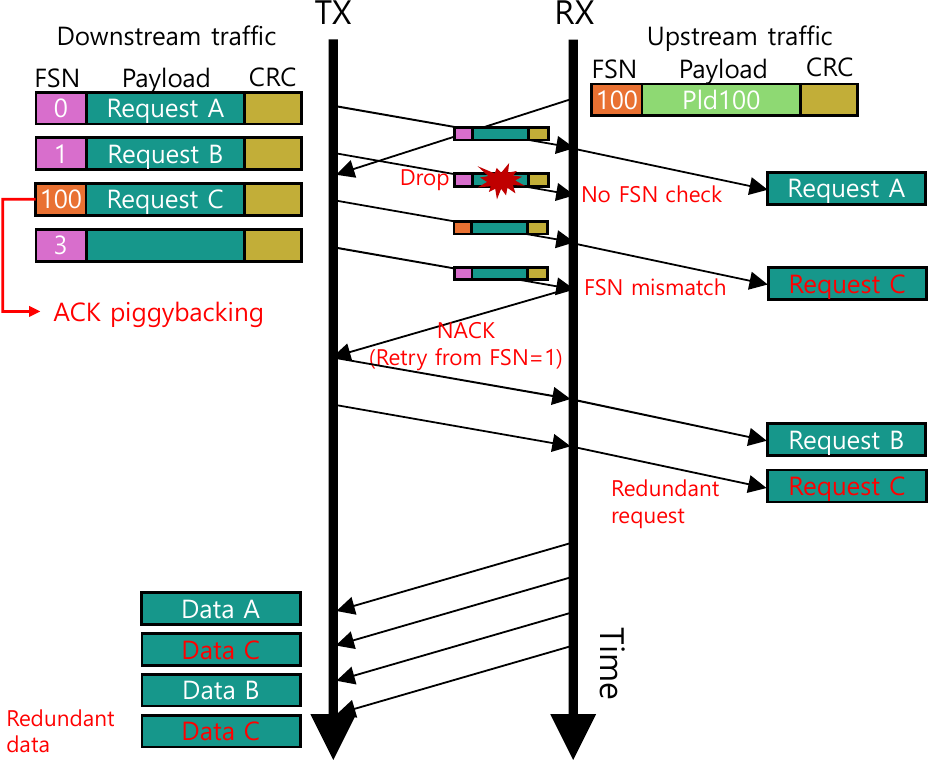}
        \label{fig:256B-failure1}
    }
    
    \subfloat[A failure due to out-of-order data.]{
        \includegraphics[scale=0.48]{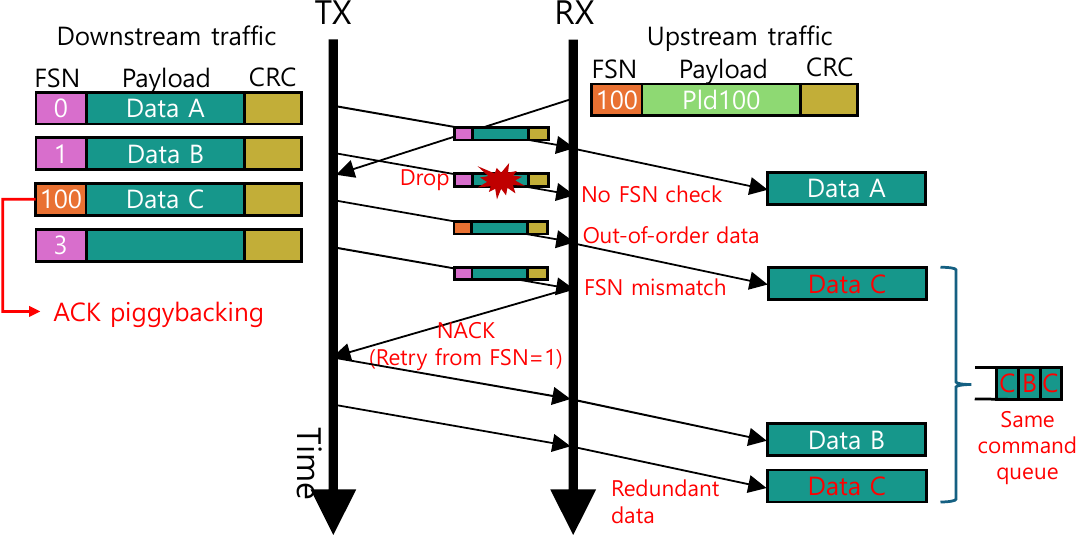}
        \label{fig:256B-failure2}
    }
    \vspace{-10pt}
    \caption{Examples of CXL's failure at the transaction layer.}
    \label{fig:unused}
    \vspace{-15pt}
\end{figure}

The reliability gap at the link layer can lead to severe consequences at higher layers, including the transaction and application layers.

Fig.~\ref{fig:256B-failure1} illustrates an example scenario involving request-carrying flits. In this example, flit \#1 (the dropped flit) and flit \#2 (the mis-forwarded flit) carry read requests B and C, respectively.  Due to the silent drop of flit \#1 and the absence of proper sequence tracking, the RX prematurely forwards flit \#2 to the upper layer, which then processes request C. Later, upon receiving flit \#3, the RX detects the missing flit and initiates a retry starting from flit \#1.

However, since request C has already been processed, the retransmitted flit leads to duplicated execution of request C. It is important to note that duplicate flit detection is typically confined to the link layer, and once a duplicate escapes this layer, there is no further protection at higher layers to detect or mitigate its effects.
As a result, the transmitter receives four data—A, C, B, and C—despite having issued only three original requests. This inconsistency may lead to FIFO overflows, buffer corruption, and misalignment between requests and responses at the transaction layer.

Fig.~\ref{fig:256B-failure2} presents another failure scenario involving data-carrying flits. CXL supports out-of-order delivery for transactions with different \emph{Command Queue IDs (CQIDs)}; that is, data associated with distinct CQIDs may arrive in any order without violating protocol correctness.
However, when multiple transactions share the same CQID, CXL requires in-order delivery of data to ensure that data is correctly reconstructed at the receiver.

In this example, data flits B and C are part of the same CQID and must be delivered sequentially. Due to the absence of proper sequence enforcement, the RX erroneously forwards flit C ahead of flit B, resulting in out-of-order delivery within the same queue. This violation may cause the application layer to process incorrect or misaligned data, compromising functional correctness.

\section{Implicit Sequence Number}
\label{sec:main}

\begin{figure}[t]
% t: top, b: bottom, h: here
    \captionsetup{width=0.5\linewidth}
    \centering
    \subfloat[CRC encoding without \mytitle{}.]{
        \includegraphics[scale=0.46]{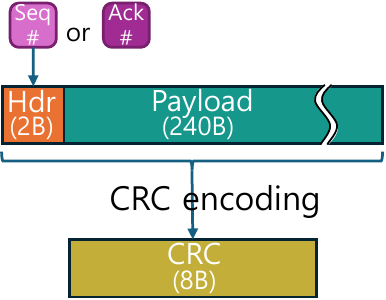}
        \vspace{-15pt}
        \label{fig:encoding1}
    }
    \hspace{30pt}
    \subfloat[CRC encoding with \mytitle{}.]{
        \includegraphics[scale=0.46]{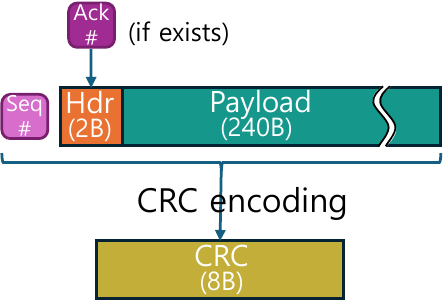}
        \vspace{-15pt}
        \label{fig:encoding2}
    }
    
    \captionsetup{width=\linewidth}
    \subfloat[An example of flit drop detection using ISN.]{
        \includegraphics[scale=0.46]{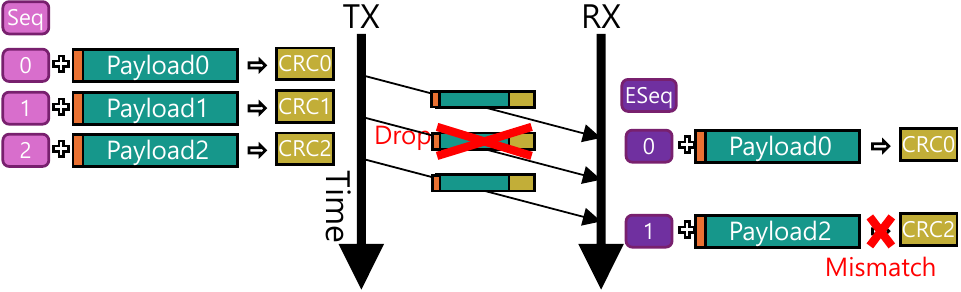}
        \label{fig:isn}
    }
    \vspace{-5pt}
    
    \caption{An overview of Implicit Sequence Number (\mytitle{}).}
    
    \vspace{-15pt}
\end{figure}

The previous section underscores the challenges of achieving both high reliability and high efficiency in CXL and other modern interconnect protocols.
To address these conflicting requirements, this paper introduces the concept of \emph{Implicit Sequence Number (\mytitle{})}, which enables accurate flit sequence tracking without explicitly transferring sequence numbers. It eliminates the need for additional header fields, maintaining high bandwidth efficiency while ensuring reliable and ordered data transmission.

The core idea of \mytitle{} is to integrate sequence numbers directly into the CRC encoding and decoding process. Similar to the baseline CXL protocol, both the sender and receiver maintain local sequence counters: \seqnum{} for the sender and \eseqnum{} for the receiver.
In the baseline CXL protocol, the sender embeds either \seqnum{} or \acknum{} in the header and generates an 8-byte CRC checksum over the 2-byte header and 240-byte payload (Fig. \ref{fig:encoding1}). As a result, only one number—either \seqnum{} or \acknum{}—can be transmitted to the opposite side.

In contrast, \mytitle{} uses the header field to carry only the \acknum{}. In cases where no acknowledgment is required (i.e., a non-piggybacking flit), this field can be filled with zeros. Meanwhile, the \seqnum{} information is not explicitly transferred within the flit, but the sender generates the CRC checksum over the header, payload, and the \seqnum{} (Fig. \ref{fig:encoding2}). The encoding process uses the same CRC polynomial over the wider input, incorporating the sequence number to provide an additional layer of sequence validation.

Figure~\ref{fig:isn} presents an example of flit drop detection using \mytitle{}. After encoding, only the payload and CRC are transmitted; the \seqnum{} itself is not included in the flit. The sender then increments its local \seqnum{} in preparation for the next transmission.
On the receiver side, the CRC is re-generated using the received 240-byte payload and the receiver's 10-bit local \eseqnum{}. If the re-generated CRC matches the received CRC, it indicates that: (1) the payload has not been altered, and (2) the sequence numbers align correctly. In this case, the receiver forwards the flit and increments \eseqnum{} for the next flit.

In the case of a dropped flit, such as $flit_{N}$, the following flit ($flit_{N+1}$) will have a CRC generated by the sender using sequence number $(N+1)$. The receiver, however, will attempt to decode it with the expected sequence number $N$ in its local counter, resulting in a CRC mismatch. This mismatch signals a flit corruption or drop, enabling the receiver to detect and handle the error accordingly.
This implicit sequence tracking ensures that any missing or out-of-order flits are quickly identified through CRC validation, thereby preserving sequence continuity without explicit sequence number transmission.

A limitation of this approach is that it verifies sequence integrity only in a binary manner (i.e., pass or fail) and does not support reordering, as it lacks explicit sequence numbers. In conventional software-centric protocols like TCP/IP, explicit sequence numbers enable the receiver to reorder out-of-order packets and ensure in-order delivery. This is typically achieved by buffering packets in system memory and allowing the operating system to reconstruct the correct sequence before delivering the data to the application~\cite{rfc815}. Such reassembly supports selective repeat—where only the missing packets are retried—and enables multi-path routing, allowing packets from the same flow to traverse different network paths to avoid congestion.

In contrast, ISN cannot support reordering because it does not transmit sequence numbers explicitly, and CRC mismatches alone cannot distinguish between flit corruption and flit drop. However, reordering is rarely implemented in high-bandwidth chip interconnects such as CXL, primarily due to the substantial on-chip storage overhead it entails. These hardware-facing protocols must deliver data directly to hardware IP blocks, precluding software-based reassembly. Supporting reordering would require dedicated on-chip reassembly buffers at the receiver to temporarily store out-of-order flits until the missing one arrives.

The required size of the reassembly buffer scales with both the link bandwidth and the worst-case skew in flit arrival times. For example, in a multi-path routing scenario with a 1 ms arrival skew, a 16-lane CXL 3.0 link operating at 1 Tbps would require a 1 Gb (128 MB) reassembly buffer. Due to such substantial overheads, the CXL 3.0 specification avoids support for multi-path routing~\cite{CXL_3_spec, cxl_intro, aurelia}.

In selective repeat scenarios, the buffer requirement is smaller but still non-trivial. For instance, if the system incurs a worst-case delay of 1 $\mu$s between flit transmission and the transmitter halting further transmission upon receiving a NACK, the receiver must provision a 1 Mb buffer to absorb in-flight flits during that window. While this is far more practical than multi-path buffering—and supported in the CXL specification—the performance benefit may be limited.
Strong FEC makes retransmissions rare, and the traffic reduction achieved through selective repeat is often marginal compared to the associated buffering cost.
As a result, many high-speed interconnects—including PCIe and CXL—favor simpler go-back-N schemes over selective repeat. In this context, ISN represents a practical and justifiable design trade-off, aligning with the cost, complexity, and performance constraints of modern chip interconnects.

\section{\mytitletwo{}: An Extension of CXL  with \mytitle{}}
\label{sec:i-seq}

Though conceptually straightforward, implementing \mytitle{} in chip interconnects like CXL requires careful consideration. This section introduces \mytitletwo{}, an extension of the CXL protocol, to deliver end-to-end protection against flit errors with minimal overhead.

\subsection{Overview}
\label{sec:i-seq:overview}

\begin{figure}[t] % t: top, b: bottom, h: here
    \centering
    \subfloat[The original CXL 3.0 protocol stack.]{
        \includegraphics[width=.98\columnwidth]{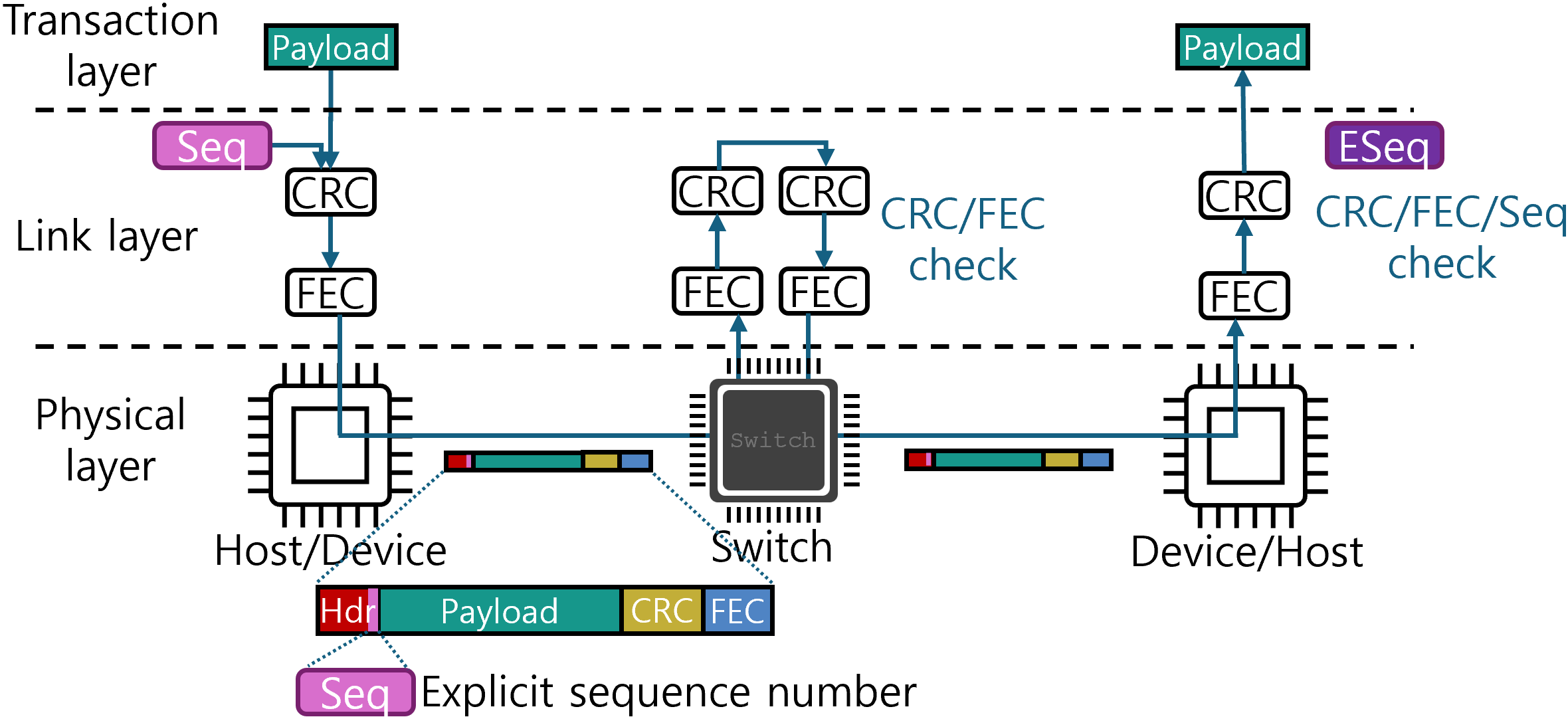}
        \label{fig:cxl-protocol-stack}
        \vspace{-12pt}
    }\\
    \centering
    \subfloat[The \mytitletwo{} protocol stack.]{
        \includegraphics[width=.98\columnwidth]{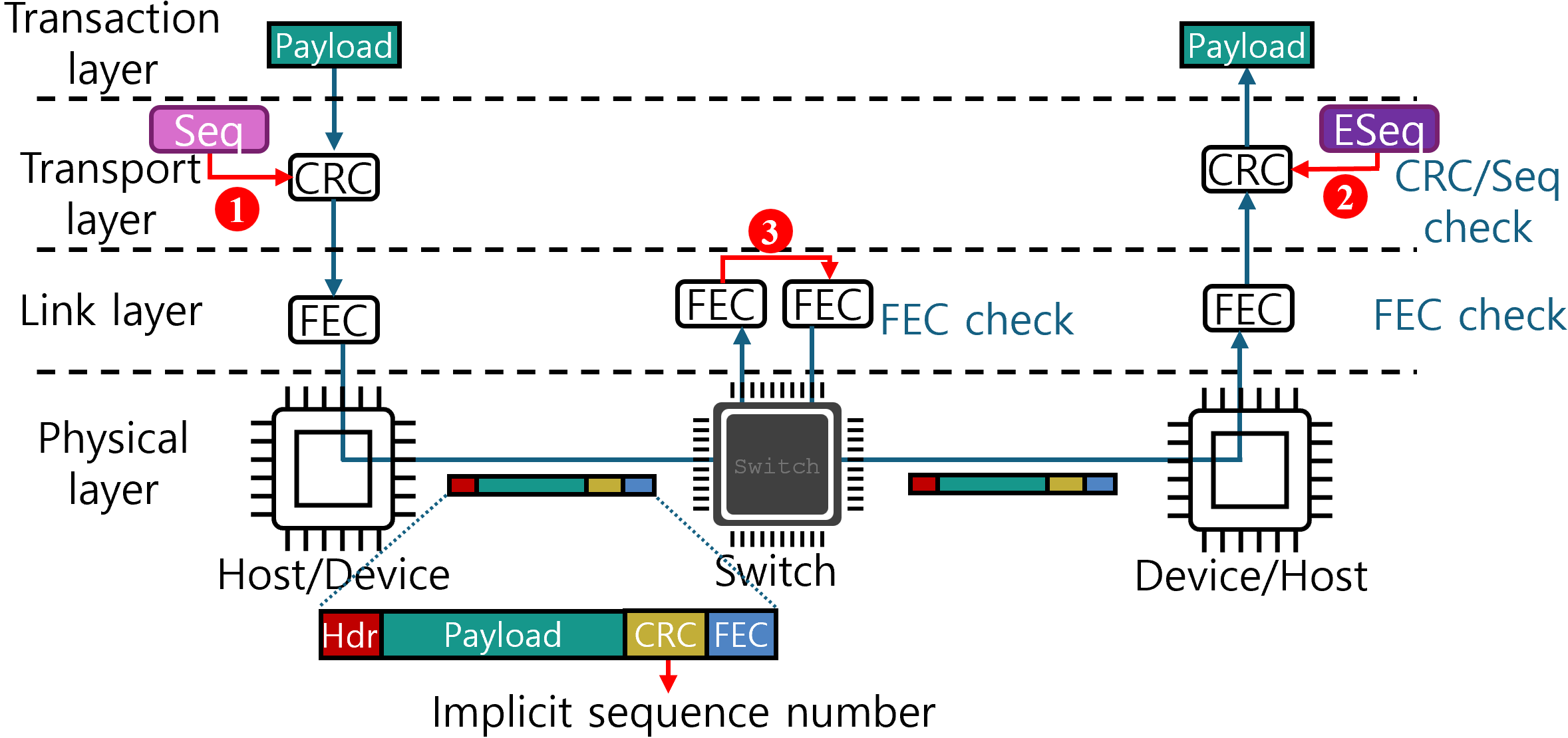}
        \label{fig:cxl+-protocol-stack}
        \vspace{-12pt}
    }\\
    \caption{A comparison of CXL and \mytitletwo{} protocol stacks.}
    \label{fig:protocol-stacks}
    \vspace{-15pt}
\end{figure}

The main challenge in implementing \mytitle{} lies in its reliance on sequence numbers for CRC decoding. This requires switching devices to maintain state and track sequence numbers of all connections passing through the switch. However, most switching devices are stateless, meaning they do not retain information about requests or responses after forwarding them. Consequently, implementing \mytitle{} in a straightforward manner may introduce additional overheads to switches, including connection state tracking and storage for sequence numbers.

To address this, \mytitletwo{} leverages the existing capabilities of FEC at the link layer. While FEC is primarily designed for error correction, it can also detect most uncorrectable errors (Section \ref{sec:background:shortened}).
Although FEC's error detection capability is not exhaustive, it provides early identification of uncorrectable errors, reducing unnecessary forwarding of erroneous flits. Any erroneous flits that bypass FEC's detection are comprehensively validated at the endpoint using a robust 64-bit CRC, ensuring an exceedingly low probability of undetected data corruption.

Therefore, \mytitletwo{} restructures the CXL protocol stack, which employs CRC and FEC at the link layer. In the revised stack (Fig. \ref{fig:protocol-stacks}), FEC remains at the link layer, correcting the majority of errors and detecting most of the uncorrectable errors. Meanwhile, CRC functionality is elevated to the transport layer, where it operates as ECRC between the endpoints. This ECRC ensures in-order delivery of flits using \mytitle{} and provides highly robust detection of flit corruption using a 64-bit CRC, enhancing both reliability and scalability.

\subsection{Flit Structure}
\label{sec:i-seq:flit}

The flit structure in \mytitletwo{} retains the original format: a 2B header with FSN and ReplayCmd fields, an 8B CRC, and a 6B FEC. The FSN field is filled with zeros in non-piggybacking scenarios (ReplayCmd = 0) and carries \acknum{} in piggybacking scenarios (ReplayCmd $\neq$ 0). Meanwhile, \mytitletwo{} ensures sequence integrity by embedding the sequence number into the CRC, highlighted as \tikz[baseline=(char.base)]{
    \node[shape=circle, draw=red, fill=red, text=white, inner sep=1pt] (char) {1};} in Fig. ~\ref{fig:cxl+-protocol-stack}. This mechanism validates sequence continuity without relying on explicit sequence numbers in the header.

To further optimize bandwidth, an alternative flit structure may omit the FSN and ReplayCmd fields entirely, reallocating their 12 bits to expand the payload or other control fields. 
In such configurations, ACK and NAK information could be conveyed as standalone packed messages rather than being embedded in the flit header (i.e., no ACK piggybacking), improving data transfer efficiency while simplifying header design. %This design mirrors the current CXL credit return mechanism, enabling efficient transfer of control information while minimizing header overhead and maximizing payload utilization.

\subsection{Endpoints}
\label{sec:i-seq:endpoints}

In \mytitletwo{}, the originator incorporates the sequence number (\seqnum{}) into the CRC generation process, ensuring that the CRC encodes both the payload and its sequence information. This CRC, combined with the payload, is then processed through FEC using the same mechanism as the original CXL protocol.
At the final destination, the FEC decoder corrects errors in the payload and CRC. The receiver validates the decoded payload and CRC against its local \eseqnum{}, as shown with \tikz[baseline=(char.base)]{
    \node[shape=circle, draw=red, fill=red, text=white, inner sep=1pt] (char) {2};} in Fig. ~\ref{fig:cxl+-protocol-stack}.

Notably, this CRC validation now resides at the transport layer (i.e., ECRC) and is performed exclusively by the endpoints. This ensures sequence continuity validation while simplifying the switches, which no longer need to track sequence numbers.
Another advantage of this transport-layer ECRC is its ability to detect errors occurring within switching devices, such as buffer corruption or switching logic errors. While link-layer CRC and FEC can protect flits during transmission across links, they do not cover errors introduced during internal operations within switches. As a result, switching devices often rely on parity or ECC for internal error protection; however, these mechanisms may leave gaps, such as undetected multi-bit errors or failures in parity protection logic. \mytitletwo{} complements these measures by ensuring that any corruption introduced within the switch is identified by the strong ECRC at the endpoint, providing robust end-to-end data integrity.

\subsection{Switches}
\label{sec:i-seq:switches}

Switches in \mytitletwo{} operate at the link layer, handling errors through FEC only (See \tikz[baseline=(char.base)]{
    \node[shape=circle, draw=red, fill=red, text=white, inner sep=1pt] (char) {3};} in Fig. ~\ref{fig:cxl+-protocol-stack}). They decode incoming flits using FEC, correcting errors up to 3-byte bursts with 3-way interleaved SSC, and re-encode the flits with FEC before forwarding them to the next node. This link-layer FEC effectively corrects most errors and detects the majority of longer burst errors.

In cases where an error is detected but uncorrectable, the switch discards the flit and reports the issue to the originator for diagnostic purposes, as in the original protocol. By limiting error handling to FEC and avoiding sequence tracking, \mytitletwo{} simplifies switch operations, reducing overhead and complexity while maintaining reliable data forwarding.

\begin{comment}
Compared to existing protocols that leverage CRC and FEC for link-layer protection, \mytitle2{}
, outlined in a step-by-step approach. We begin with a basic configuration where a host and CXL device are directly connected, communicating via 68B flits. From this foundation, we expand the configuration to cover more complex and general scenarios, ultimately leading to a robust, complete solution for CXL 3.0.

\subsection{Direct Communication in 68B Flits}
\label{sec:i-seq:config1}
In this configuration, TX

\subsection{Switched Communication in 68B Flits}
\label{sec:i-seq:config2}
In this configuration,

\subsection{Switched Communication in 256B Flits}
\label{sec:i-seq:config3}
In the third configuration, 

\end{comment}

\section{Evaluation}
\label{sec:eval}

This section evaluates the reliability and performance impacts of \mytitletwo{} compared to the existing CXL protocol, alongside an analysis of its hardware overhead.

\subsection{Reliability}
\label{sec:eval:reliability}

We define a protocol failure as 1) a flit with corrupted data being forwarded to the application layer ($Fail_{data}$), or 2) a flit being forwarded to the application layer in an incorrect order ($Fail_{order}$).
The application-level impact of these failures depends on the flit's contents and application behavior, which are beyond the scope of this evaluation. Instead, this analysis focuses on assessing the occurrence rates of these protocol failures. The evaluation begins with configurations involving direct connections and extends to configurations involving multi-level switching.

\subsubsection{CXL in Direct Connections}
\label{sec:eval:256B_DC}
In this configuration, the host and device are directly connected via a single CXL link, and we assume that all flits arrive at the endpoint.
We calculate the \emph{Flit Error Rate (FER)} based on independent bit errors. Assuming a BER of $10^{-6}$ and a flit\_size of 2048 bits (256 bytes) in CXL 3.0:
\begin{equation}
    FER = 1-(1-BER)^{flit\_size} \simeq 2.0 \times 10^{-3}
\end{equation}
This means approximately $0.2\%$ of flits are erroneous, equating to 1 million erroneous flits out of 500 million flits per second on a $\times16$ CXL link.

Next, we assess the probabilities of these erroneous flits being uncorrectable and detectable by FEC and CRC. The lower bound for uncorrectable flit error rate after FEC is provided by the PCIe 6.0 standard~\cite{PCIe_6_spec}:
\begin{equation}
    FER_{UC} = 3.0 \times 10^{-5}
\end{equation}

From this, we can expect FEC to correct more than 98.5\% of erroneous flits, as derived from:
\begin{equation}
    p_{correct} = 1 - \frac{FER_{UC}}{FER}
\end{equation}

Among the remaining $1.5\%$ uncorrectable flits, the 64-bit CRC detects the vast majority with an exceedingly low undetected error probability of $2^{-64}$. The resulting undetectable flit error rate is
\begin{equation}
    FER_{UD} = FER_{UC} \times 2^{-64} \simeq 1.6 \times 10^{-24}
\end{equation}
Note that this $FER_{UD}$ represents an upper bound, as burst errors no longer than 64 bits are detected with 100\% certainty.

Given a $\times16$ CXL device transferring 500M flits per second, this undetected error rate translates to a \emph{Failure In Time} (FIT, number of failures expected over one billion hours) of:
\begin{equation}
    FIT_{device} = FER_{UD} \times 500,000,000 \times 3,600 \times 10^{9} \simeq 2.9 \times 10^{-3}
\end{equation}

This exceptionally low FIT value demonstrates that the direct connection configuration provides extremely high reliability, surpassing the typical target for server-grade devices, which have FIT values in the range of a few hundred~\cite{Server_FIT}.

\subsubsection{CXL in Single-level Switched Environments}
\label{sec:eval:CXL_switch}

In this configuration, the host and device communicate through a single-level switch, with data transmitted over two links.
The switch discards uncorrectable flits detected on the first link between the originator and the switch. As a result, the rate of dropped flits at the endpoint is the same as the rate of uncorrectable flits at the first link:
\begin{equation}
    FER_{drop} = FER_{UC} = 3.0 \times 10^{-5}
\end{equation}

Dropped flits can lead to ordering failures when the subsequent flit conveys an \acknum{} instead of a \seqnum{}. The probability of this occurring depends on the ACK coalescing level, denoted as 
$p_{coalescing}$.
For example, $p_{coalescing}=0.1$ indicates one in ten flits carries an \acknum{}.
The ordering failure rate (with $p_{coalescing} = 0.1$) is thus given by:
\begin{equation}
    FER_{order} = FER_{drop} \times p_{coalecing} = 3.0 \times 10^{-6}
\end{equation}

This rate far exceeds the undetected data failure rate after CRC and is used as the primary consideration for simplicity. For a $\times16$ device transmitting 500M flits per second, the resulting FIT is:
\begin{equation}
    FIT_{device} = FER_{order} \times 500,000,000 \times 3,600 \times 10^{9} \simeq 5.4 \times 10^{15}
\end{equation}

This extraordinarily high FIT value of ordering failures highlights the vulnerability of the current CXL protocol in switched environments. The strong detection and correction mechanisms of CRC and FEC, while effective for data integrity, do not address this ordering failure mode, as they operate exclusively on received flits and do not mitigate flit drops.
In CXL, flit drops are detected exclusively through explicit sequence numbers. However, the use of ACK piggybacking can prevent some flits from carrying their sequence numbers, leaving these flits unprotected against drops and resulting in undetected ordering violations.

\subsubsection{\mytitletwo{} in Single-level Switched Environments}
\label{sec:eval:CXL+_switch}

This analysis considers the same interconnect topology as before (single-level switching) but upgrades the protocol to \mytitletwo{}. Unlike the original CXL protocol, \mytitletwo{} integrates \mytitle{}, allowing it to detect all flit drops and prevent ordering failures ($Fail_{order}$) through retries. Consequently, the only remaining failure type is corrupted data being forwarded to the application layer ($Fail_{data}$).

Flits dropped on the first link are eventually identified by the endpoint with \mytitle{} and retried. The probability of data failures at the endpoint, after accounting for retries, is given by:
\begin{equation}
FER_{UD} = (1 + FER_{UC}) \times 2^{-64} \simeq 1.6 \times 10^{-24}
\end{equation}
This extremely low probability is driven by the 64-bit CRC’s strong error detection capability.

The corresponding FIT rate for a $\times16$ device is:
\begin{equation}
FIT_{device} = 1.6 \times 10^{-24} \times 500,000,000 \times 3,600 \times 10^9 \simeq 2.9 \times 10^{-3}
\end{equation}

This reduction in FIT value, exceeding $10^{18}$ times lower than that of the original CXL protocol, demonstrates the dramatic reliability improvements achievable by \mytitletwo{}. By eliminating ordering failures and ensuring robust end-to-end error detection, \mytitletwo{} provides an efficient and reliable solution for scaling out chip interconnect networks in multi-node systems.

\subsubsection{Multi-level Switched Environments}
\begin{figure}[t]
    \centering
    \includegraphics[width=0.9\columnwidth]{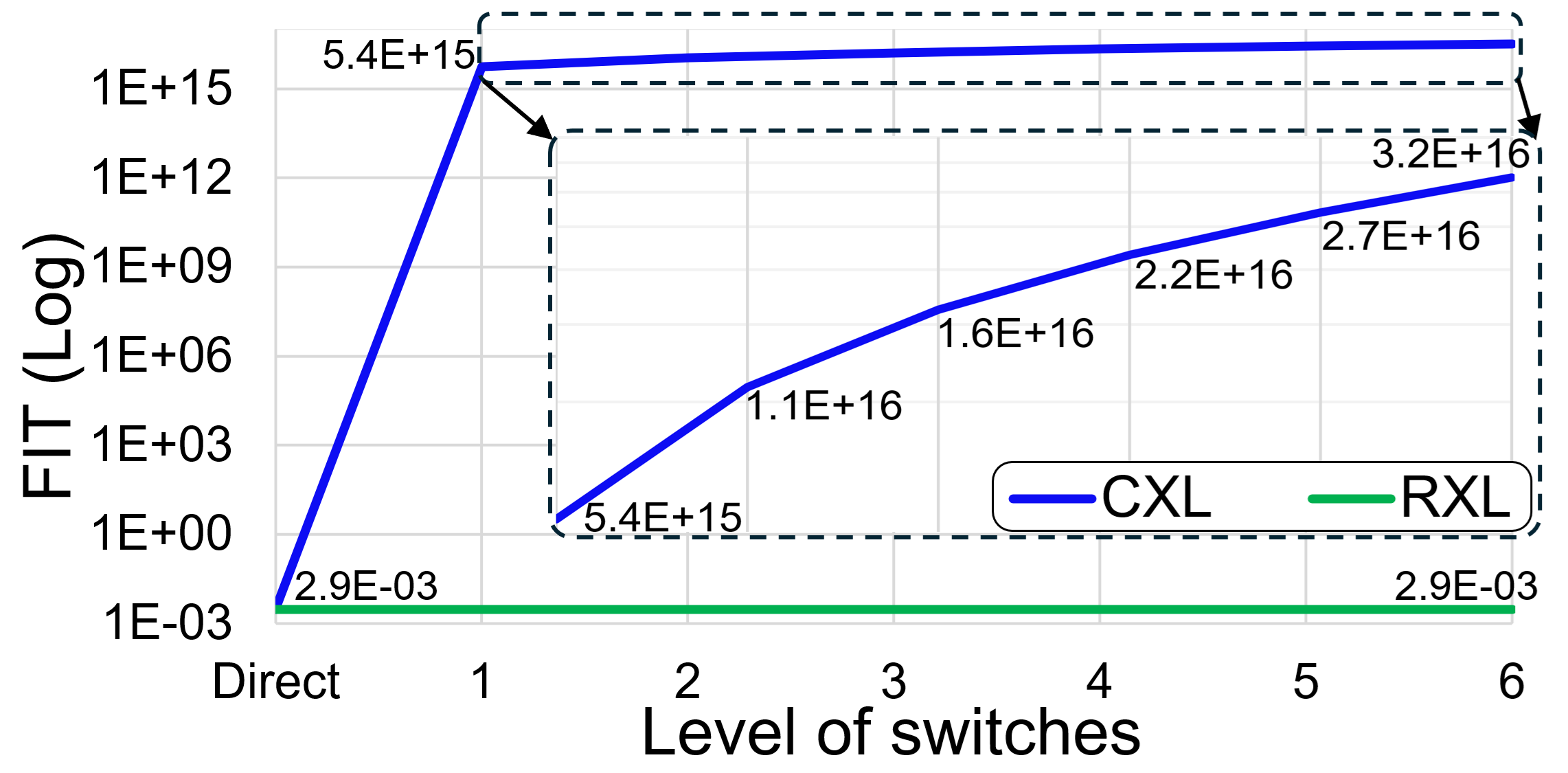}
    \vspace{-5pt}
    \caption{Comparison of $FIT_{device}$ of CXL and \mytitletwo{} against increasing switching levels.}
    \vspace{-10pt}
    \label{fig:FIT_per_Hop}
\end{figure}

The dropped flits and ordering failures in CXL pose significant challenges to scaling out systems, especially in multi-level switched environments. Fig. \ref{fig:FIT_per_Hop} compares the $FIT_{device}$ of CXL and \mytitletwo{} as the level of switching increases.
Both protocols exhibit high reliability in direct connection configurations. However, with the introduction of single-level switching, the reliability of CXL declines drastically, decreasing by approximately $10^{18}$ times due to its susceptibility to flit drops and sequence violations. Furthermore, these failure rates apply to a single device; in multi-node systems with numerous parallel connections, the compounded effect makes the current CXL protocol insufficient for maintaining reliable chip interconnect networks.

In contrast, the reliability of \mytitletwo{} remains nearly unchanged, even with increasing levels of switching. Its embedded sequence validation mechanism ensures robust error detection and ordering integrity, effectively mitigating vulnerabilities that undermine CXL in switched environments.
This analysis underscores the importance of incorporating advanced sequence tracking mechanisms like \mytitletwo{} to maintain reliability and efficiency in switched and large-scale interconnect environments.

\subsection{Performance}
\label{sec:eval:performance}

The performance impact of the protection schemes is evaluated by estimating the bandwidth loss they introduce. This analysis considers a $\times16$ link transmitting 256B flits every 2ns. A go-back-N retry scheme with a 100ns latency is assumed~\cite{PCIe_numbers}, during which the link is occupied with retransmitted flits.

\subsubsection{CXL in Direct Connections}

Bandwidth loss due to retries is calculated by comparing the effective duration of flit transmission with and without retries. Flits without a retry, occurring with a probability of $(1 - FER_{UC})$, occupy the channel for 2ns each. Conversely, flits requiring retries, occurring with a probability of $FER_{UC}$, occupy the channel for 102ns. The bandwidth loss can thus be estimated as:
\begin{equation}
    \label{eqn:bwloss1}
    BW_{loss} = 1 - \frac{2ns}{(1-FER_{UC})\times2ns + FER_{UC} \times 102ns} \simeq 0.0015
\end{equation}

This calculation shows that retries result in a bandwidth loss of approximately 0.15\%, demonstrating that CXL ensures reliable and efficient communication even under retry scenarios in direct connections. Moreover, adopting a more efficient retry scheme, such as selective retry, could further reduce this loss, enhancing overall link utilization and system performance.

\subsubsection{CXL in Switched Environments}

In this analysis, we explore two options for CXL to handle flit drops.
The first option continues to use ACK piggybacking, minimizing performance degradation by avoiding the overhead of separate ACK flits. However, this approach is vulnerable to frequent protocol failures caused by flit drops.

In this scenario, flits dropped at the first link are eventually detected by the explicit sequence number of the immediately following or subsequent flits at the endpoint, initiating retries. The rate of such dropped flits at the first link is equal to $FER_{UC}$. Similarly, uncorrectable flit errors at the second link, also occurring at a rate of $FER_{UC}$, initiate additional retries. Consequently, the cumulative rate of flits requiring retries is $2 \times FER_{UC}$. The bandwidth loss from these retries can be estimated similarly to Eqn.~\ref{eqn:bwloss1}:
\begin{equation}
    \label{eqn:bwloss2}
    BW_{loss} = 1 - \frac{2ns}{(1-2\times FER_{UC})\times2ns + 2\times FER_{UC} \times 102ns} \simeq 0.0030
\end{equation}

This calculation indicates that the first option results in a small bandwidth loss of approximately 0.3\%. However, this approach suffers from a critical reliability issue related to flit drops, as it utilizes piggybacked ACKs and cannot detect some flit ordering failures.

The second option disables ACK piggybacking and instead uses separate flits for ACK and NACK messages. This approach ensures precise sequence tracking by embedding explicit sequence numbers in each flit. However, it incurs a significant bandwidth cost due to the additional overhead of separate ACK flits.

The bandwidth loss from these ACK flits can be estimated as:
\begin{equation}
    \label{eqn:bwloss3}
    BW_{loss} = p_{coalescing}
\end{equation}
Without ACK coalescing ($p_{coalescing} = 1$), the bandwidth loss could reach 100\% as the link is fully utilized to transfer ACK flits in the reverse direction. To mitigate this overhead, coalescing more ACK messages into a single ACK flit (reducing $p_{coalescing}$) can be employed. 
However, achieving higher levels of ACK coalescing requires larger buffers to store transmitted flits until the corresponding ACK is received. This increased buffering demand adds system complexity and resource requirements, and it also imposes practical limits on the coalescing level.

\subsubsection{\mytitletwo{} in Switched Environments}

With \mytitletwo{}, all flits dropped at the first link, occurring at a rate of $FER_{UC}$, are immediately detected by their subsequent flits through embedded sequence tracking, initiating a retry. Similar to the CXL approach, almost all uncorrectable flit errors at the second link are detected by the ECRC and trigger a retry. As a result, the rate of retries is nearly identical to that calculated in Eqn.~\ref{eqn:bwloss2}:
\begin{equation}
    \label{eqn:bwloss4}
    BW_{loss} = 1 - \frac{2ns}{(1-2\times FER_{UC})\times2ns + 2\times FER_{UC} \times 102ns} \simeq 0.0030
\end{equation}

This result demonstrates that \mytitletwo{} incurs similar bandwidth loss (approximately 0.3\%) as the CXL approach with ACK piggybacking while providing enhanced reliability by embedding sequence validation directly into the ECRC.

This evaluation highlights that \mytitletwo{} ensures robust data and sequence integrity without adversely affecting performance, even in multi-level switching environments. By seamlessly integrating sequence validation into the error-checking mechanism, \mytitletwo{} addresses the shortcomings of CXL, such as vulnerability to sequence violations. As a result, \mytitletwo{} provides a scalable and efficient solution tailored to the demands of modern high-speed interconnects, offering enhanced reliability while maintaining high performance.

\subsection{Hardware Overheads}

Although \mytitle{} integrates \seqnum{} into CRC encoding and decoding, it introduces minimal hardware overhead. In the encoding process, the 10-bit \seqnum{} is XORed with the lower 10 bits of the 240B payload. The XORed result then feeds into the existing CRC encoder to generate the checksum according to the new scheme, adding only 10 parallel XOR gates.
Similarly, in CRC decoding, the \eseqnum{} is XORed with the received payload before entering the existing decoder. 
As a result, both the encoder and decoder require only 10 XOR gates, with an increase of just one additional logic depth. Additionally, ISN eliminates the need for the 10-bit comparator that previously compared \seqnum{} and \eseqnum{}. Consequently, the overall hardware overhead from ISN is minimal, requiring only a few additional gates.

\section{Conclusion}
\label{sec:conclusion}

This paper presents \mytitle{}, a novel concept to achieve reliable and efficient chip interconnect, and \mytitletwo{}, a scalable and reliable extension of the CXL protocol.
By embedding sequence tracking into CRC, \mytitle{} eliminates the need for explicit sequence numbers in flit headers, reducing header overhead without compromising sequence integrity. The proposed \mytitletwo{} enhances the robustness of the chip interconnect by ensuring end-to-end data and sequence validation while maintaining compatibility with existing flit structures.

Our evaluation demonstrates that \mytitletwo{} achieves the same level of performance as the existing method while addressing the critical reliability vulnerability of ordering failures.
Additionally, \mytitletwo{} delivers strong end-to-end protection against data corruption, ensuring even errors internal to switches are effectively detected and mitigated by 64-bit CRC.
These features make \mytitletwo{} a highly effective solution for modern high-speed, error-prone interconnect environments.

The lightweight nature of \mytitle{} and its seamless integration with existing protocols position it as a scalable and efficient approach for chip interconnects. Future work could explore extending \mytitle{} to other protocols and systems, such as Network-on-Chip (NoC) and chiplet interconnects, or adapting its principles to address challenges in broader communication protocols, further enhancing reliability across diverse computing architectures.

\bibliographystyle{ACM-Reference-Format}
\bibliography{refs}

\end{document}